# Evaluating Collaborative Search Interfaces with Information Seeking Theory

Max L. Wilson, m.c. schraefel
School of Electronics and Computer Science
University of Southampton, UK
{mlw05r, mc}@ecs.soton.ac.uk

**ABSTRACT**

Despite the many implicit references to the social aspects of search within Information Seeking and Retrieval research, there has been relatively little work that has specifically investigated the additional requirements for collaborative search software. In this paper we re-assess a recent evaluation framework, designed for individual information seeking experiences, to see a) how it could still be applied to collaborative search software; b) how it could produce additional requirements for collaborative search; and c) how it could be extended in future work to be even more appropriate for collaborative search evaluation. The position held after the assessment is that it can be used to evaluate collaborative search software, while providing new insights into their requirements. Finally, future work will validate the frameworks applicability to collaborative search and investigate roles within collaborative groups as a means to extend the framework.

**ACM Classification Keywords**
H.5.2 User Interfaces: Evaluation/methodology, Prototyping. H.5.3 Group and Organization Interfaces: Computer-supported cooperative work. H.1.2 User/Machine Systems: Human factors.

**INTRODUCTION**

Although forms of implicit collaboration, such as recommender systems, have been well researched, investigation into interfaces for explicit, synchronous and asynchronous, collaborative search software has only recently received a flurry of interest. This is surprising given that such collaboration has been identified many times in the history of information seeking research, discussed in detail by Hansen and Järvelin [4], and there has been around 20 years of research into Computer Supported Collaborative Work (CSCW). The focus for Collaborative Information Retrieval (and Seeking) research, however, is a union of these two areas that extends our ideas of Information Seeking research with a subset of the tasks being investigated by CSCW.

Given the relatively infant level of investigation so far, into teams or groups who search together to achieve a mutual goal, recent efforts have focused on identifying the additional requirements for collaborative search software. An example is the survey performed by Morris [5], that revealed that around 95% of people have performed in weekly or monthly collaborative searches. The investigation further reveals the methods used by people when achieving their collective goal.

In this paper we continue to identify requirements for collaborative search software by assessing a recent evaluation framework. The framework was developed using models of information seeking behavior, which have historically focused on individuals. The assessment is designed to see a) how it could still be applied to collaborative search; b) what additional requirements it could identify for collaborative search software; and c) how it may be extended in the future to be more appropriate for collaborative search conditions. The content of the paper should also provide the additional means required for researchers to apply the framework to collaborative search software, which we intend to do in our own future work for validation purposes.

**RELATED WORK**

As opposed to collaborative search environments that use techniques such as relevance feedback, which have been deemed as implicit [8], explicit collaborative information retrieval involves groups of people actively participating as a team to gather information on a shared goal. Further Pickens and Golovchinsky [8] break this down into synchronous and asynchronous, where groups are either working together in real-time or are collaborating without any immediate social communications, respectively.

Recent efforts have produced some early designs of explicit collaborative search software that, in turn, are also producing new insights into additional requirements for collaboration during information seeking tasks. $S^3$, standing for Storable, Shareable Search [6], was designed to support asynchronous explicit search by recording peoples searching activities, making them persistent over time, and providing them to others in a team. Synchronous explicit collaboration can be further broken down into co-located and distributed groups. An example of co-located collaborative search software is CoSearch [9], which lets groups of people use a combination of mobile devices to search as a group over one machine. An example of a distributed explicit synchronous application is SearchTogether [7], which provides means of communicating with, recommending pages to and monitoring the activity of other searchers. SearchTogether does, however, allow for these communications to be persistent and so can work for asynchronous groups too.

**EVALUATION FRAMEWORK**

Recent work by Wilson et al. [11, 12] has produced an evaluation framework that is designed to systematically inspect prototype interfaces in terms of the tactics they allow users to employ and the types of conditions the searchers may be in. Consequently, it can tell evaluators a) which types of users are well or poorly supported, b) which types of tactics users may struggle to perform, and c) which parts of the interface are enabling the tactics that are supported. These tactics are discussed in more detail below, but examples include broadening a search, checking what has been done, and weighing up options. User types are broken down by dimensions such as their existing knowledge and confidence in finding an answer. This approach, based in theory, was also recently validated [10].

To perform such a detailed and systematic inspection of prototype designs, the framework relies on information seeking theory. Specifically, the framework uses one model of tactics, by Bates [1, 2], and one model of users, by Belkin et al. [3]. One contribution of the framework was to provide a connection between the two models that states which tactics are most useful for each user type. Below, we assess these two models from information seeking theory, to see how they can be applied to collaborative search.

**Bates' Model of Tactics**

Bates identified 32 different tactics that people may carry out when searching for information across different technologies. Where these were originally designed to be self-serving tactics, they may have different implications for those who are part of a group or team. We now step through these tactics to identify the additional considerations that evaluators must maintain when applying the framework to collaborative search software.

The first five tactics are 'Monitoring Tactics'.
- CHECK is to check that the current state of search is still related to the original reason for searching. In a group setting, the user may have to check both their current task, and the overall task of the group.
- WEIGH is to consider whether to continue or choose a different approach. In a group setting, users will require knowledge of what approaches have already been tried by other members of the team.
- PATTERN is to monitor ones actions for efficiency. In a group setting, users may benefit from comparing their own patterns to those of co-searchers.
- CORRECT involves watching for and correcting any errors during search. Although this may maintain as an individual activity, the many eyes of others may help identify errors a user has missed. Thus, in a group setting, it may be helpful to notice errors in other peoples work.
- RECORD is to record items for later return. The capture of context here may be even more important for others in the group who did not perform the original search.

The following 7 tactics relate to parsing result sets.
- BIBBLE is to check to see if other searchers have already carried out the current task. This may change vary little, except that those who may have already carried out the work may be others in the team, rather than unknown searchers from the past.
- SELECT is to select part of a task and address it as a set of sub-tasks. In a group setting, it may be beneficial to know that others have not already completed these sub-tasks, or to see if others could share the workload.
- SURVEY is to review the current available options. Again, it may be of value to know that others have not already completed some of current options.
- CUT is to take an action that has the largest affect on the overall task. This may not vary in collaborative search software, as other tactics from this group deal with preparing for the decision.
- STRETCH is similar to reusing something. It may be that a user can 'stretch the value' of someone else's hard work to benefit their own. The actions of a known team of group may be much easier to visualize than trying to browse the previous actions of every other user in the history of the search service.
- SCAFFOLD is to design a different approach to find a certain result, having followed a 'dead end' path. This may be much easier to do if the user can see and mimic the successful paths taken to similar targets by others.
- CLEAVE is a fairly solo activity in terms of applying a binary search technique to going through a structured list.

The following 6 tactics relate to formulating search plans, which has been shown as a core activity during collaborative search [5].
- SPECIFY is to apply a set of query terms that are known to produce the desired result. Searchers may benefit from knowledge from others in the group to do this, especially those who are not search-savvy.
- Being EXHAUSTive is also an activity that is easier with a team of searchers.
- To REDUCE is the opposite of EXHAUST, which allows un-expected but potentially valuable results to be found. This often involves parsing a larger amount of results, with many being unrelated or previously found and so shared human resources may help here too.
- PARALLEL is to broaden a search by using synonymous terms, for example. Like EXHAUST, this may be easier with shared group knowledge.
- To PINPOINT is the opposite of PARALLEL, and allows for searching to focus on specific synonyms.
- BLOCK relates, for example, to the use of 'NOT' in a Boolean query. In a group, this action may help avoid overlap and may help searchers to discover results on a certain topic, but avoid results that relate to what a colleague is searching for.

The next 11 tactics relate to the specific terms used after having formulated a search plan: SUPER, SUB, RELATE, NEIGHBOR, TRACE, VARY, FIX, REARRANGE, CONTRARY, RESPELL and RESPACE. We do not discuss these individually here, but they are each mainly solo decisions. They could still benefit, however, from an awareness of others peoples search terms and phrases.

The final 3 tactics relate to changing ideas or mental concepts of the searcher and so tend to relate to the on-going learning that informs better searching behavior. Consequently, the three tactics are important for a team setting for keeping each other informed and sharing specific advances on a goal or problem.

- RESCUE is to rethink a problem, when the searcher realizes their ideas are inherently incorrect.
- BREACH is to extend ones boundaries of understanding given new information. An example may be realizing that diabetes is not solely related to genetics, but also to aspects such as diet.
- FOCUS, therefore, is the opposite of BREACH and relates to identifying that only a sub-part of a problem is actually relevant to the overall goal.

It is clear that collaborative search interfaces should support the transfer of developed understanding to other members of a team, as they may significantly alter the direction of the whole group.

Most of the discussion of tactics above could be generalized to the need to either actively share results or passively monitor the progress of others. In the evaluation framework, however, each feature of an interface, such as the keyword search form, the list of results, the communication channels, and so on, are addressed individually in terms of how they support each tactic. This process, therefore, means that the evaluator is encouraged to think about how the keyword search box, for example, can also be used to indicate to the user that a search has been carried out before by another group member. Consequently, it leads to a system where support of tactics is pervasive to the whole interface rather than having specific functions or features of the design that specifically support individual tactics. This should become clearer as we discuss a specific example below.

**Belkin et al's Model of Users**
Unlike Bates' model of tactics, there are no changes or additional considerations to the model by Belkin et al, when applying the framework to collaborative software. The dimensions of the model are: Method (scanning or actively searching), Goal (to learn or to select and take away), Mode (by recognizing or being able to specify), and Resource (looking for a report, for example, or information about a report). None of these dimensions are related to or affected by the need to collaborate with others. Instead they reflect the individual's activity when performing their tasks for the benefit of the team. We revisit the idea of types of users in our future work section below, but first we discuss an example application of the framework.

**EVALUATING A COLLABORATIVE SEARCH UI**
The framework is designed to evaluate specific interfaces to a system. So to understand the CoSearch system [9], for example, which allows people to take part in the search using devices such as mobile cellular phones, the evaluator would consider how both the mobile interface and the computer interface support users. The combination of interfaces makes CoSearch a complex example. Instead, SearchTogether makes a good clear example, as it provides a single interface for every user in the team, and has been specifically designed for collaborative web searching [7].

Unfortunately there is not the available space in this workshop paper for a full evaluation, like the ones previously published by Wilson et al. [11, 12]. Instead we discuss some of the specific features of the software that have been developed. It is important to note that in order to perform a full evaluation, familiarity with the interface is always beneficial. Having novice users of a particular system perform the evaluation, however, can provide insights into discoverability of interface features.

The first step of applying the evaluation is to identify each of the features in the interface, such as the keyword search box and the results list. In the case of SearchTogether, there is also the ability to do a split search and a multiple-engine search, to comment and recommend, chat via instant messaging, a complete search summary, a view of individual search histories and the ability to view the pages that are currently being browsed by other collaborators.

Each of the identified features of the interface will have the ability to support many of the 32 tactics. Step 2, therefore, is to measure the minimum number of moves it may take to perform each of the tactics with each of the identified features. For more information on this unit of measurement, refer to earlier publications [11, 12]. The output of these two steps, which could be applied to multiple user interfaces if doing a comparison, is a series of graphs that tell you a) the breadth of support each interface feature is producing, b) the amount to which each tactic is being supported, and then c) which types of users are, therefore, being well supported.

Let us consider the group query history function, for example, which shows the user search phrases already used by each of the other team members. This directly supports many tactics from the list described above, especially as it can remain persistently in view throughout the user's search sessions. For it example, the query history function helps users to identify PATTERNs; to BIBBLE and make sure that no one has performed a valuable search before; to SURVEY their options; STRETCH other peoples searches; and many more in terms of devising specific queries.

The group history function above, however, is an example of one that has been specifically designed to support collaborative search. It is also possible for keyword searches to help provide this similar context of other's queries. For example, when a user enters a search phrase, the interface could indicate that, if applicable, someone else has performed it before. This would support the user in avoiding overlap, without having to proactively double-check each query.

The search summary view is another important feature of the interface, but depending on how it is designed, it could support different tactics. In this case, the user can use it to view previously found pages. This could help the user TRACE the pages for useful search terms, or may simply help with the final 3 idea tactics. If the search summary also recorded the search terms used to find the annotated item, users could both BIBBLE and SCAFFOLD their own follow-up searches. Similarly, knowing the search terms that found a poorly rated page could help the user BLOCK certain terms.

The above examples show how the framework can provide both insights into the support being provided and identify potential design ideas for improving the design. In the final example of the search summary, by simply asking how the feature could also support the BIBBLE tactic, new ideas can be produced.

**FUTURE WORK**

Despite the fact that Belkin's model is unchanged by the notion of collaboration, there are different types of searchers in teams and each of them may well benefit from different types of support. Pickens and Golovchinsky [8], for example, identify two team dynamics in their paper: domain experts working together, and domain experts working with search experts (where the latter may be a librarian for example). Similarly, in the scenarios provided to inform the SearchTogether interface [7], Morris and Horvitz also identify different roles in the groups. The elderly father, for example, may benefit more from the search experience of his son or daughter, by receiving recommendations, than by being able to see what they have each searched for and found in the past. The search-experienced son, though, may benefit more by seeing a quick overview of what everyone has found so far, to be able to consider new search directions.

Our position is that this notion of roles may be important for understanding the value of, and therefore in evaluating, collaborative search software. A similar example could be seen when a manager may be assigning tasks to employees. For the manager, summaries and overviews may be valuable, but for the employees, avoiding duplication of effort and having effective means of sharing results may be more important. Consequently, part of our future work will involve investigation into CSCW research to understand what is known about roles within teams. The aim is to identify an additional model to the framework to be used when evaluating collaborative search software.

**CONCLUSIONS**

In this paper we have analyzed an evaluation framework for information seeking interfaces in terms of its applicability to collaborative search software. The investigation indicates that the framework can be just as easily applied to collaborative search interactions as individual seeking software, but that there are additional considerations about the individual's involvement within a group that must be maintained as the assessment is carried out. These additional considerations for each tactic are listed above, so that researchers have the additional means required to apply the framework to collaborative search interfaces. Our own future work will include applying the framework to a collaborative search interface to validate the conclusions of this paper. Other future work will also investigate including a model of team roles as a means to enhance the framework for collaborative search interfaces.